\documentclass[]{article}
\usepackage{pdfpages} 
\usepackage{lmodern}
\usepackage{amssymb,amsmath}
\usepackage{ifxetex,ifluatex}
\usepackage{fixltx2e} % provides \textsubscript
\ifnum 0\ifxetex 1\fi\ifluatex 1\fi=0 % if pdftex
  \usepackage[T1]{fontenc}
  \usepackage[utf8]{inputenc}
\else % if luatex or xelatex
  \ifxetex
    \usepackage{mathspec}
  \else
    \usepackage{fontspec}
  \fi
  \defaultfontfeatures{Ligatures=TeX,Scale=MatchLowercase}
\fi
\IfFileExists{upquote.sty}{\usepackage{upquote}}{}
\IfFileExists{microtype.sty}{%
\usepackage[]{microtype}
\UseMicrotypeSet[protrusion]{basicmath} % disable protrusion for tt fonts
}{}
\PassOptionsToPackage{hyphens}{url} % url is loaded by hyperref
\usepackage[unicode=true]{hyperref}
\hypersetup{
            pdftitle={Automatic differential analysis of NMR experiments in complex samples},
            pdfauthor={Laure Margueritte1; Petar Markov2; Lionel Chiron3; Jean-Philippe Starck3; Catherine Vonthron-Sénécheau1; Mélanie Bourjot1; Marc-André Delsuc4*},
            pdfborder={0 0 0},
            breaklinks=true}
\usepackage{longtable,booktabs}
\IfFileExists{footnote.sty}{\usepackage{footnote}\makesavenoteenv{long table}}{}
\usepackage{graphicx,grffile}
\makeatletter
\def\maxwidth{\ifdim\Gin@nat@width>\linewidth\linewidth\else\Gin@nat@width\fi}
\def\maxheight{\ifdim\Gin@nat@height>\textheight\textheight\else\Gin@nat@height\fi}
\makeatother
\setkeys{Gin}{width=\maxwidth,height=\maxheight,keepaspectratio}
\IfFileExists{parskip.sty}{%
\usepackage{parskip}
}{% else
\setlength{\parindent}{0pt}
\setlength{\parskip}{6pt plus 2pt minus 1pt}
}
\providecommand{\tightlist}{%
  \setlength{\itemsep}{0pt}\setlength{\parskip}{0pt}}
\ifx\paragraph\undefined\else
\let\oldparagraph\paragraph
\renewcommand{\paragraph}[1]{\oldparagraph{#1}\mbox{}}
\fi
\ifx\subparagraph\undefined\else
\let\oldsubparagraph\subparagraph
\renewcommand{\subparagraph}[1]{\oldsubparagraph{#1}\mbox{}}
\fi

\makeatletter
\def\fps@figure{htbp}
\makeatother

\title{Automatic differential analysis of NMR experiments in complex samples}
\author{Laure Margueritte\textsuperscript{1} \and Petar Markov\textsuperscript{2} \and Lionel Chiron\textsuperscript{3} \and Jean-Philippe Starck\textsuperscript{3} \and Catherine Vonthron-Sénécheau\textsuperscript{1} \and Mélanie Bourjot\textsuperscript{1} \and Marc-André Delsuc\textsuperscript{4}*}
\providecommand{\institute}[1]{}
\institute{Université de Strasbourg, CNRS, Laboratoire d'Innovation Thérapeutique
(LIT), UMR7200, Labex MEDALIS, 67000, Strasbourg, France \and Structural Biophysics Group, School of Optometry and Vision Sciences,
Cardiff University, United Kingdom \and CASC4DE Le Lodge, 20, Avenue du Neuhof, 67100 Strasbourg, France \and Institut de Génétique et de Biologie Moléculaire et Cellulaire (IGBMC),
INSERM U596, CNRS UMR 7104, Université de Strasbourg,
Illkirch-Graffenstaden, France}
\date{}

\begin{document}
\maketitle

\begin{enumerate}
\def\labelenumi{\arabic{enumi}.}
\tightlist
\item
  Université de Strasbourg, CNRS, Laboratoire d'Innovation Thérapeutique
  (LIT), UMR7200, Labex MEDALIS, 67000, Strasbourg, France
\item
  Structural Biophysics Group, School of Optometry and Vision Sciences,
  Cardiff University, United Kingdom
\item
  CASC4DE Le Lodge, 20, Avenue du Neuhof, 67100 Strasbourg, France
\item
  Institut de Génétique et de Biologie Moléculaire et Cellulaire
  (IGBMC), INSERM U596, CNRS UMR 7104, Université de Strasbourg,
  Illkirch-Graffenstaden, France
\end{enumerate}

\vskip 1cm

\section{Abstract}\label{abstract}

Liquid state NMR is a powerful tool for the analysis of complex mixtures
of unknown molecules. This capacity has been used in many analytical
approaches: metabolomics, identification of active compounds in natural
extracts, characterization of species, and such studies require the
acquisition of many diverse NMR measurements on series of samples.

While acquisition can easily be performed automatically, the number of
NMR experiments involved in these studies increases very rapidly and
this data avalanche requires to resort to automatic processing and
analysis.

We present here a program that allows the autonomous, unsupervised
processing of a large corpus of 1D, 2D and DOSY experiments from a
series of samples acquired in different conditions. The program provides
all the signal processing steps, as well as peak-picking and bucketing
of 1D and 2D spectra, the program and its components are fully
available. In an experiment mimicking the search of an active species in
natural extract, we use it for the automatic detection of small amounts
of artemisin added to a series of plant extracts, and for the generation
of the spectral fingerprint of this molecules.

This program called Plasmodesma is a novel tool which should be useful
to decipher complex mixtures, particularly in the discovery of
biologically active natural products from plants extracts, but can also
in drug discovery or metabolomics studies.

\section{Introduction}\label{introduction}

Liquid state NMR is a powerful tool for the analysis of mixtures
containing unknown molecules. All species in the solution display their
NMR spectra, with a signal intensity proportional to their relative
concentrations, provided that slow tumbling rates or relaxation agents
do not hide the lines by fast relaxation processes. This capacity has
been used in many analytical approaches: metabolomics, identification of
active compounds in natural extracts, characterization of
species\textsuperscript{1--5}. See references (6), (7), and (8) for
recent reviews.

Such studies require the acquisition of many diverse NMR measurements on
series of samples. Modern NMR spectrometers allow sequential actions
(introduction of the sample, probe tuning, acquisition) in order to
produce automatically the corresponding 1D and 2D data sets.

Unfortunately, if the acquisition is quite easily performed, the access
to the final informations is less straightforward: signal processing
(Fourier transform, phasing, baseline correction, peak detection) and
finally spectrum interpretation are not trivial tasks. Moreover, the use
of 2D spectra implies more complex steps and additional tasks such as
reduction of t\textsubscript{1}-noise and t\textsubscript{1}-ridges, or
the determination of contour levels for display.

In the case of metabolomics studies, or natural extracts screening, the
number of NMR experiments increases very rapidly and this data avalanche
requires to resort to automatic processing. While metabolomics are aimed
at measuring precisely the amount of well-known compounds, and to
quantify precisely their variations from sample to sample, the
identification of an active molecule in a natural extracts starts with
its detection and then its characterization of an unknown compond or
eventually a family of related species.

In this article, we present the specific development of a computer
program allowing the autonomous, unsupervised processing of a large
corpus of 1D and 2D experiments from a series of samples acquired in
different conditions.

Results obtained using this program on series of complex natural
extracts highlight the time saving and the efficiency increase regarding
classical ``hand-made'' processing of raw data.

\section{Software developments}\label{software-developments}

\subsection{Plasmodesma}\label{plasmodesma}

The program Plasmodesma\textsuperscript{9} developped for this project
relies on the SPIKE library for most of its
operation\textsuperscript{10}. It is intended to process autonomously a
large series of different spectra originated from different samples,
obtained in varying conditions. This analytical process involves the
handling of a complex set of NMR experiments (1D and 2D homo- or
hetero-nuclear spectra), at the end, a spectral report summarizing the
analysis is expected, containing all figures, peak and bucket lists for
each sample.

The current work is based on SPIKE (\textbf{S}pectrometry
\textbf{P}rocessing \textbf{I}nnovative \textbf{KE}rnel)a comprehensive
software on which the current work is based, is a comprehensive software
library aimed at the processing and analysis of Fourier transform
spectroscopies. It provides basic functionalities such as apodization,
Fourier transforms, phasing, peak-picking, line-fitting, baseline
correction as well as more advanced tools. It is easily extensible
through a plug-in mechanism. SPIKE combines the use of a parallel multi
processor approach to a low memory footprint, thus insuring rapid
processing with an optimal use of the computer hardware. Moreover, SPIKE
allows the efficient handling and visualization of very large data-sets
limited only by disk space. SPIKE is a continuation of the previous Gifa
and NPK\textsuperscript{11,12} NMR processing softwares, and was
developed to include other Fourier transform spectroscopies, in
particular FT Mass spectrometry (Orbitrap and FT-ICR) and
2D-FT-ICR\textsuperscript{13--15}. For portability reasons and ease of
development, the program is written in Python, and relies on external
libraries such as \texttt{numpy}, \texttt{scipy}, \texttt{pandas}, and
\texttt{hdf5}.\textsuperscript{16--20}

\subsection{Principle of operations}\label{principle-of-operations}

The program Plasmodesma operates without any human interaction. When
applied to a folder, all NMR files are imported, processed, and a global
report is generated for the totality of the analysis. All the processing
and analysis steps are optimized depending on the acquisition parameters
found in the data-sets, either 1D or 2D data. No other input is
required. The 1D and 2D experiments are processed sequentially, The data
are apodized, Fourier transformed, and the baseline corrected, 1D
spectra are also automatically phased. Additionally, an efficient
denoising step\textsuperscript{21} is performed on the 2D experiments,
in order to reduce the t\textsubscript{1}-noise. The F1 Fourier
transform step of the 2D data-sets is performed depending on the
spectral type and on the acquisition protocol. The calibration is then
determined precisely from the reference signal (in this case, 0 ppm for
the TMS). A peak-picking and bucket analysis are then performed (see
below). Peak lists and bucket lists are generated as csv files for each
experiment. Finally, figures of each spectrum is created, with and
without peaks displayed. A final report that contains all acquisition
and processing parameters is generated (see S.I. S1).

\subsection{Specific developments}\label{specific-developments}

Some functions used by Plasmodesma have been developed specifically for
this analysis, and were implemented as SPIKE plug-ins.

\paragraph{Autophasing}\label{autophasing}

In the context of metabonomics and screening studies, the possibility to
detect and quantify precisely the intensity of vanishing small peaks is
paramount. The phase of a 1D spectrum, if set slightly off, may have a
strong impact on the possibility to detect small signals, in particular
if they are close to a large one. Errors of only a few degrees introduce
bias resulting to too low or too high quantization, as well as shifts of
the maximum. Automatically acquired natural extract spectra are usually
difficult to phase because of strong solvent lines and other artifacts
present in the spectra. The improvement of the simple but robust
automatic phasing procedure developed in NPK\textsuperscript{12}
contributes efficiently to resolve this problem. The principle is to
minimize the negative wing of the 1D spectrum, by performing a grid
search first on 0\textsuperscript{th} order (frequency independent)
alone, then on both 0\textsuperscript{th} and 1\textsuperscript{st}
order (frequency dependent) corrections, the larger peak being used as
the 1\textsuperscript{st} order pivot. An automatic baseline correction
(see below) is performed at each correction step, and an optional
\emph{inwater} mode allows to ignore the central spectral zone.

\subsubsection{Autobaseline}\label{autobaseline}

A flat baseline is also a requisite for correct analysis, and a specific
plugin as been developed in this respect. We developed a new approach,
which relies on an iterative statistical treatment on the signal split
into pieces of constant length, and fitting the baseline by piecewise
linear segments. The fit is based on the use of a linear regression
minimizing the \(\ell_p(x) = \left( \sum(|x|^p ) \right)^{1/p}\) norm of
the difference. A rough estimate of the spectral baseline is first
generated using \(p=1\) on each pieces. Then, the estimate is
iteratively improved by removing that part of the signal above the
current baseline approximation, and using \(p=3\) for fitting. This
method guarantees baselines that stick well to the signal avoiding
spurious oscillations that higher-order polynomials or splines may
produce.

\subsubsection{Bucketing}\label{bucketing}

Bucketing is an important operation in the processing pipeline. It
consists in computing the area under the spectrum over small spectral
segments which cover the whole spectral width. The segments should be
large enough to blur the small discrepancies that appear from one sample
to another, while preserving the resolutive power of the spectra. A
bucket size of 0.01 ppm was used for 1D \textsuperscript{1}H spectra.

Bucketing also reduces the size of the data that will be submitted to
statistical analysis. This is of foremost importance in the analysis of
2D spectra, which routinely contains millions of points. The reduction
of 2D datasets to tractable sizes in statistical tools requires
nevertheless bucket sizes on the order of 0.03 to 0.05~ppm in
\textsuperscript{1}H spectroscopy and to 1.0~ppm in
\textsuperscript{13}C. Such sizes are certainly too large to capture all
the details contained in the 2D spectra. One solution to this difficulty
could be to use segments of varying size, however we rather chose to
enrich the information by adding to the area of each bucket, additional
information. For each bucket, computed over 1D or 2D spectra, the
coordinates of the bucket center and its size in pixel were stored,
along with the area information computed as the mean over the bucket,
and enriched with the values of the min and max points, and the standard
deviation of data over the bucket.

\subsubsection{Processing of DOSY
experiment}\label{processing-of-dosy-experiment}

DOSY spectra are extremely efficient in deciphering complex mixtures,
and have been used in many different work (see \emph{Mahrous et
al}\textsuperscript{7} and reference therein). They require a specific
processing for the analysis of the exponential decays observed along the
indirect dimension of the 2D spectrum. In this work this specific
processing was performed by using the recently introduced PALMA
algorithm\textsuperscript{22} that implements a rapid Inverse Laplace
Transform analysis, using a hybrid constraint, maximizing the entropy
while minimizing the \(\ell_1\) norm of the reconstructed spectrum. This
algorithm was developed using the SPIKE library, so it was particularly
easy to insert it into the processing pipe-line. As a consequence, they
are systematically processed and a peak list and an adapted bucket list
is also generated.

\subsubsection{Report}\label{report}

Finally, a concise report is produced as a csv file (see S.I. S2). The
report contain all the important parameters related to data acquisition
and processing. They are finally displayed, as rendered using the pandas
python library\textsuperscript{18}.

\subsection{Analysis}\label{analysis}

Given a set of 1D and 2D NMR raw experiments, the approach described
above is able to produce, in full automation and without any human
interaction, a set of correctly processed spectra, along with complete
peak lists and enriched bucket lists.

The artifacts observed in the spectra, such as antidiagonals,
t\textsubscript{1} noise and ridges, etc. were corrected on the bucket
list. On modern spectrometers these artifacts are usually at a low
intensity, however as the purpose here is to detect species at low
concentration, and their presence is detrimental.

The 2D bucket list is corrected for remaining t\textsubscript{1} noise
and t\textsubscript{1} ridges for each column in the matrix, by setting
to null all buckets below twice the median value of the considered
column. The bucket list originated from symmetric spectra, such COSY and
TOCSY, were further corrected for departure of this symmetry by setting
symmetrical buckets to the minimum value of the pair. These two
operations have the effect of preserving the most significant buckets,
without loosing the weak spectral areas.

\section{M\&M}\label{mm}

\textbf{Chemicals.} Artemisinin 98\% was purchased in Sigma-Aldrich and
deuterated methanol (10 x 0.75mL) in Eurisotop (Saint Aubin, France).

\textbf{Algae collection and identification.} The algae \emph{Sargassum
muticum} was collected in June 2006 in Cap Lévy (Manche), France.
Taxonomic determination was performed by Dr A-M. Rusig and a voucher
specimen was deposited in the Herbarium of the University of Caen.
Extraction was realized as in Vonthron-Sénécheau et
al.\textsuperscript{23}.

\textbf{Samples preparation.} Five samples containing 10 mg of \emph{S.
muticum} hydroalcoholic extract were prepared. An artemisinin DMSO
solution at 3 mg/mL was added to the samples to obtain a final
concentration of artemisinin of 0.2, 0.3, 0.4 and 2.7 mg/ml in NMR
tubes, as summarized in Table 1. All samples were lyophilized and
dissolved in 750~µL of methanol d4, and put in 5~mm NMR tubes. The NMR
tubes were spun with a small bench centrifuge to help sedimentation of
insoluble parts, and placed in the NMR sample changer.

\begin{longtable}[]{@{}rccccc@{}}
\caption{Samples preparation}\tabularnewline
\toprule
Sample n° & 1 & 2 & 3 & 4 & 5\tabularnewline
\midrule
\endfirsthead
\toprule
Sample n° & 1 & 2 & 3 & 4 & 5\tabularnewline
\midrule
\endhead
\emph{S. muticum} extract & 10 mg & 10 mg & 10 mg & 10 mg & 10
mg\tabularnewline
added artemisinin & 0 mg & 0.15 mg & 0.24 mg & 0.32 mg & 2
mg\tabularnewline
\bottomrule
\end{longtable}

A sample of pure artemisinin was prepared in methanol d4 and studied by
NMR.

\subsubsection{NMR spectroscopy}\label{nmr-spectroscopy}

\textbf{Acquisitions} were performed on a Bruker Avance-III spectrometer
operating at 700~MHz, and equipped with a TCI cryo probe and a standard
Bac60 sample changer. Each sample was automatically inserted into the
spectrometer, tuned and shimmed after a stabilization delay of 120
seconds. All experiments were automatically run on each sample, the
whole sequence being programmed using a TopSpin macro (see E.S.I S2).
Spectral parameters (\(\pi / 2\) pulses, receiver gain\ldots{}) were
optimized on one sample and used for the whole series without further
check.

Spectral widthes were set to 12~ppm in \textsuperscript{1}H and to
150~ppm in \textsuperscript{13}C. 1D spectra were acquired on 64 scans,
16384 points, and a relaxation delay of 1.5~sec, for a total time of 3
minutes. COSY experiments were performed with 8 scans, with 512
increments of 4096 points each, for a total acquisition time of 2 hours.
TOCSY experiments were performed with 8 scans, with 400 increments of
4096 points each, and using a DISPSI-2 mixing sequence of 80~msec
duration. TOCSY acquisition time was 1 hour 40 minutes. DOSY experiments
were performed with 32 scans, with 50 increments of 4096 points each,
for a total acquisition time of 50 minutes. HSQC experiments were
performed with 4 scans, with 512 increments of 2048 points each, for a
total acquisition time of 1 hour. HMBC experiments were performed with
48 scans, with 400 increments of 4096 points each, for a total
acquisition time of 10 hours.

The complete acquisition time for one sample, including sample injection
and tuning, took about 16 hours. The five samples were acquired in one
continuous run.

The artemisinin sample was studied by NMR: Artemisinin:
\textsuperscript{1}H NMR (CD3OD, 700 MHz) \(\delta\) 0.99 (3H, d,
\emph{J} = 6.2 Hz, 6-CH3), 1.16 (3H, d, \emph{J} = 7.2 Hz, 9-CH3), 1.38
(3H, s, 3-CH3), 2.08 (1H, ddd, H4), 2.40 (1H, ddd, H4), 2.01 (1H, m,
H5), 1.47 (1H, m, H5), 1.38 (1H, m, H5a), 1.52 (1H, m, H6), 1.09 (1H, m,
H7), 1.77 (1H, m, H7), 1.17 (1H, m, H8), 1.86 (1H, m, H8), 1.82 (1H, m,
H8a), 3.31 (1H, dq, H9), 6.03 (1H, dq, H12), \textsuperscript{13}C NMR
(CD3OD, 700 MHz) \(\delta\) 106.7 (C, C3), 25.2 (CH3, C3), 36.6 (CH2,
C4), 25.7 (CH2, C5), 51.2 (CH, C5a), 38.1 (CH, C6), 19.9 (CH3, C6), 34.6
(CH2, C7), 24.0 (CH2, C8), 45.6 (CH, C8a), 34.0 (CH, C9), 12.7 (CH3,
C9), 81.0 (CH, C12a), 95.5 (CH, C12).

\subsubsection{Data Processing}\label{data-processing}

\textbf{Spectral Processing} was integrally performed using the
Plasmodesma program presented here. The program is written in python,
and is compatible both with python 2 and python 3. It is based on the
SPIKE library\textsuperscript{10} and the DOSY processing were performed
using the PALMA approach\textsuperscript{22} embedded in SPIKE as a
plugin.

Complete processing took 96 minutes on a MacOs machine, running the
python anaconda distribution 4.2 from Continuum Analytics (Austin, TX).

\textbf{Statistical Analysis.} The bucket lists and peak lists produced
by the Plasmodesma run were analyzed with a python script based on the
pandas library, using the Jupyter notebook environment.

The Plasmodesma program, along with examples, experimental data related
to the artemisin series, Jupyter notebooks presenting the data analysis,
and the specific SPIKE plugins are freely available at
https://github.com/delsuc/plasmodesma repository.

\section{Results}\label{results}

To mimic the presence of a bioactive molecule at different
concentrations in complex mixtures, crude plant extracts were
supplemented at different concentrations with artemisinin, a naturally
occurring and structurally known sesquiterpene lactone, and five
different samples were prepared. All five samples were placed in the
sample changer and NMR data were acquired in an automatic manner, after
an initial tuning of the first sample.

\subsection{Data Processing}\label{data-processing-1}

\begin{figure}
\centering
\includegraphics[width=0.55000\textwidth]{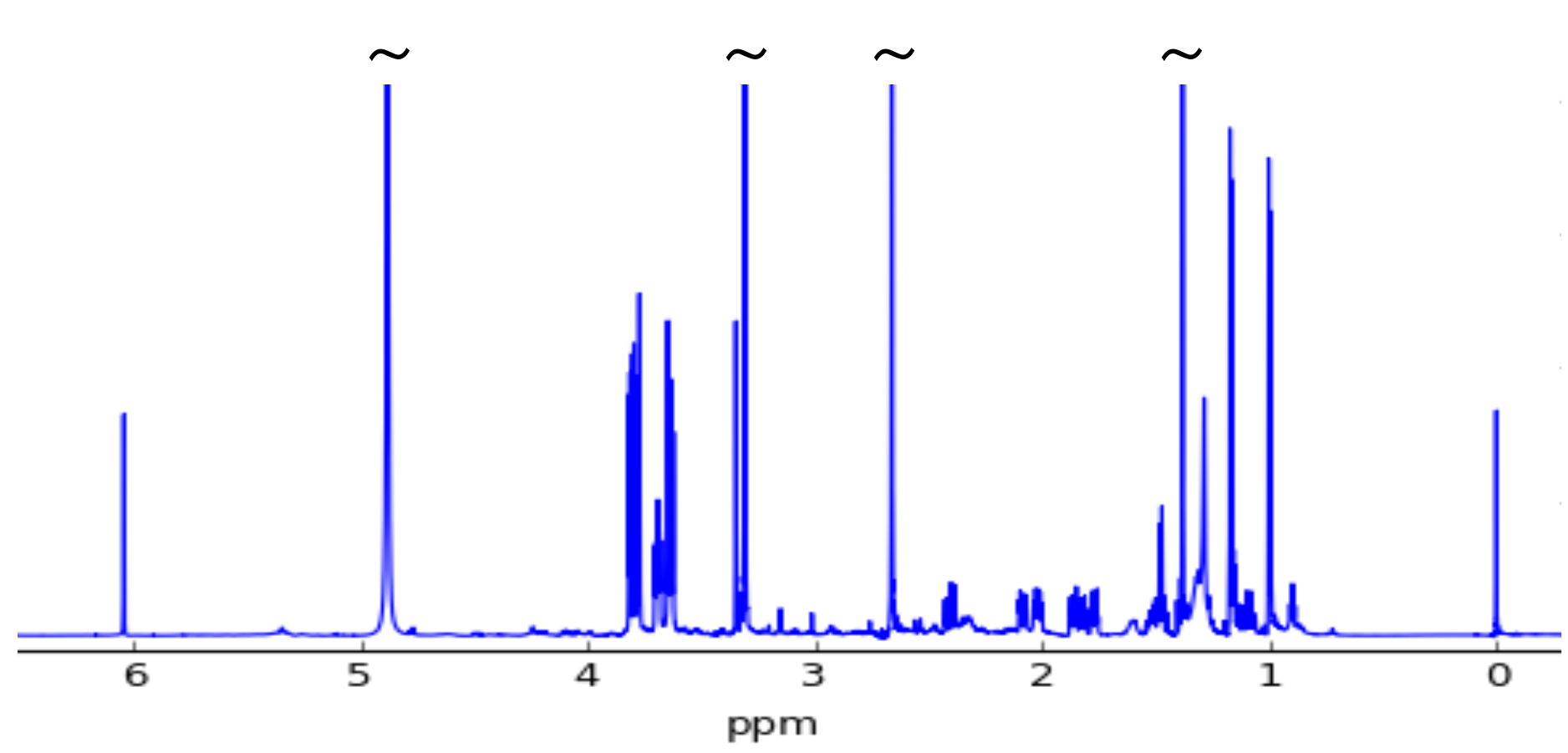}
\caption{1D \textsuperscript{1}H spectrum of sample n°5.\label{fig:1D}}
\end{figure}

\begin{figure}
\centering
\includegraphics{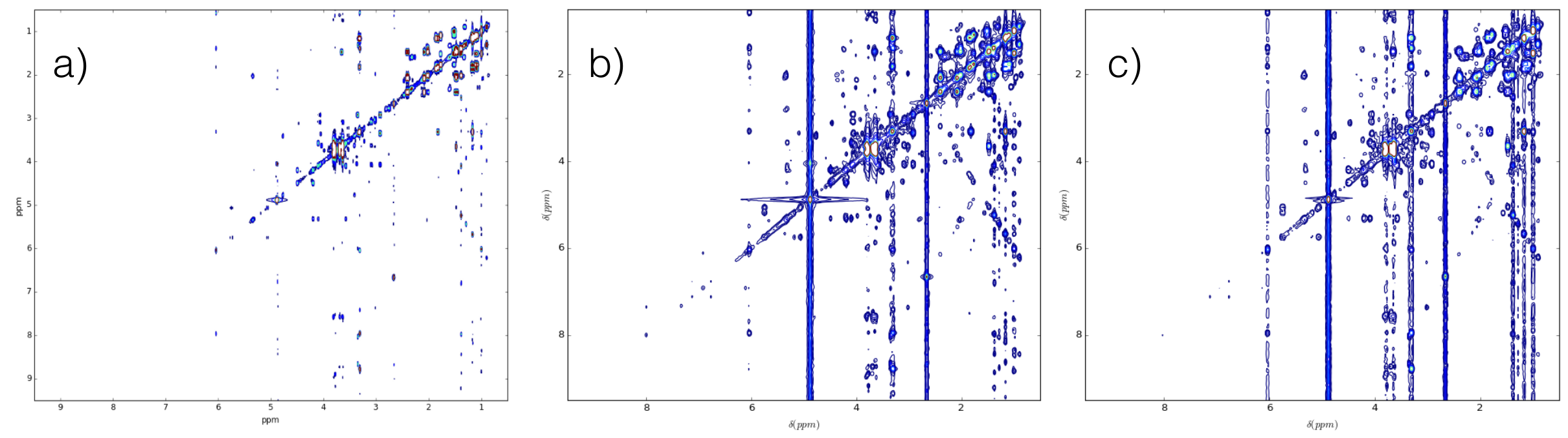}
\caption{COSY spectrum of sample n°5. a) automatically generated contour
plot, b) graphic representation of the raw bucket list, c) same as b,
but showing the standard deviation of each bucket.\label{fig:COSY}}
\end{figure}

The raw data-sets were processed as described above, and the peak lists
and bucket lists generated. Figures \ref{fig:1D} and \ref{fig:COSY} show
an example of the result of such a processing.

The bucketing procedure is used to summarize the spectral content, and
by reducing the size of the data to handle, to ease further statistical
analysis. However, it can be seen in Figure \ref{fig:COSY} that
t\textsubscript{1}-noise and other spectral artifacts are present in
particular in the standard deviation analysis, which enhances the local
signal variations. It appears that corruption of buckets from spectral
artifacts appear more deleterious in 2D spectroscopy than in classical
1D. For this reason, the areas and standard deviation values of the
bucket list were subjected to the simple procedures described above. The
first step consists in setting to a null value all values below a
certain threshold computed from the median over the vertical column of
the considered bucket. This procedure allows to remove a large part of
the noise, and to only retain the peaks separated above the threshold.
The threshold level adapted for each column permit to efficiently clean
the strong t\textsubscript{1}-noise stripes, while preserving weak peaks
located in less crowded regions. In a second step, homonuclear
experiments a symmetrization procedure can be applied, it was done here
by simply taking the smaller of the two values related by symmetry. This
procedure is much simpler and more robust on bucket lists than on real
spectra, as the bucketing has already homogenized the spectral axes and
produced squared buckets. This procedure was applied on the area and the
standard deviation values of the bucket list.

\begin{figure}
\centering
\includegraphics{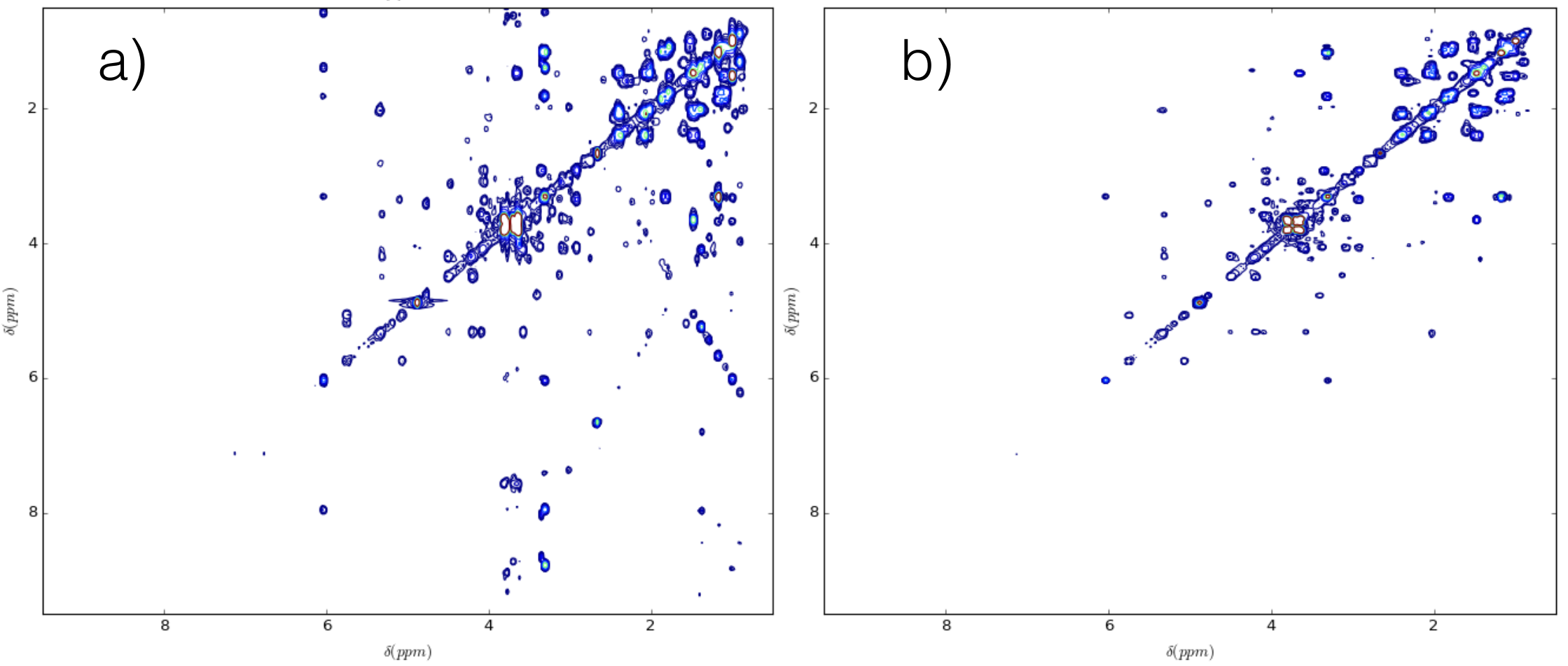}
\caption{a) Standard deviation the bucket list from Figure
\ref{fig:COSY} after t\textsubscript{1}-noise removal, b) same data as
in a after symmetrization.\label{fig:COSYclean}}
\end{figure}

Figure \ref{fig:COSYclean} shows the result of each cleaning steps. It
can clearly be seen that this procedure, allows an improvement of the
quality of data and a better compatibility with automatic analysis.

\subsection{Data Analysis}\label{data-analysis}

\begin{figure}
\centering
\includegraphics{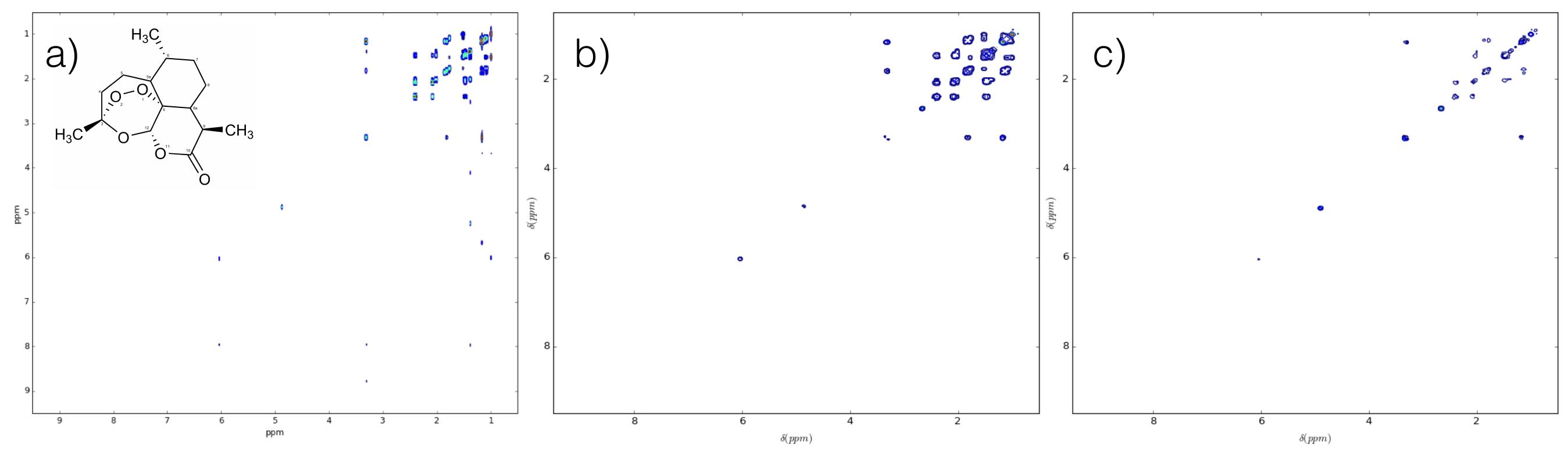}
\caption{Automatic extraction of spectral fingerprint from the automatic
spectral analysis a) standard COSY spectrum of artemisinin, b) result of
the subtraction of the std channel of the bucket lists of sample 5 and
sample 1, c) same as b for sample 4 and sample 1.\label{fig:Diff}}
\end{figure}

The cleaned bucket lists can be efficiently used for detection of the
spectral features varying from spectrum to another. This can be done on
any 1D or 2D spectra: Figure \ref{fig:Diff} shows the result on the
analysis of the COSY spectrum.

Obviously, the positions of the signals of the artemisinin spectrum is
detected and separated from the constant background, even though the
background is of much larger intensity. Here the original spectrum is
not genuinely recovered, not only because some signals are missing, but
principally because of the loss of the intensities. However, the
generated spectral pattern can be used to extract chemical shifts and
topologies, and recognize a molecular pattern, which can be used as a
fingerprint. The same result cannot be obtained directly from the
spectrum, and is efficient because the bucketing standardizes the
spectra, the standard deviation measures the fluctuation rather the
intensity. Finally the cleaning operation smooth out the random
fluctuations which otherwise would hamper the direct comparison to
operate.

\begin{figure}
\centering
\includegraphics{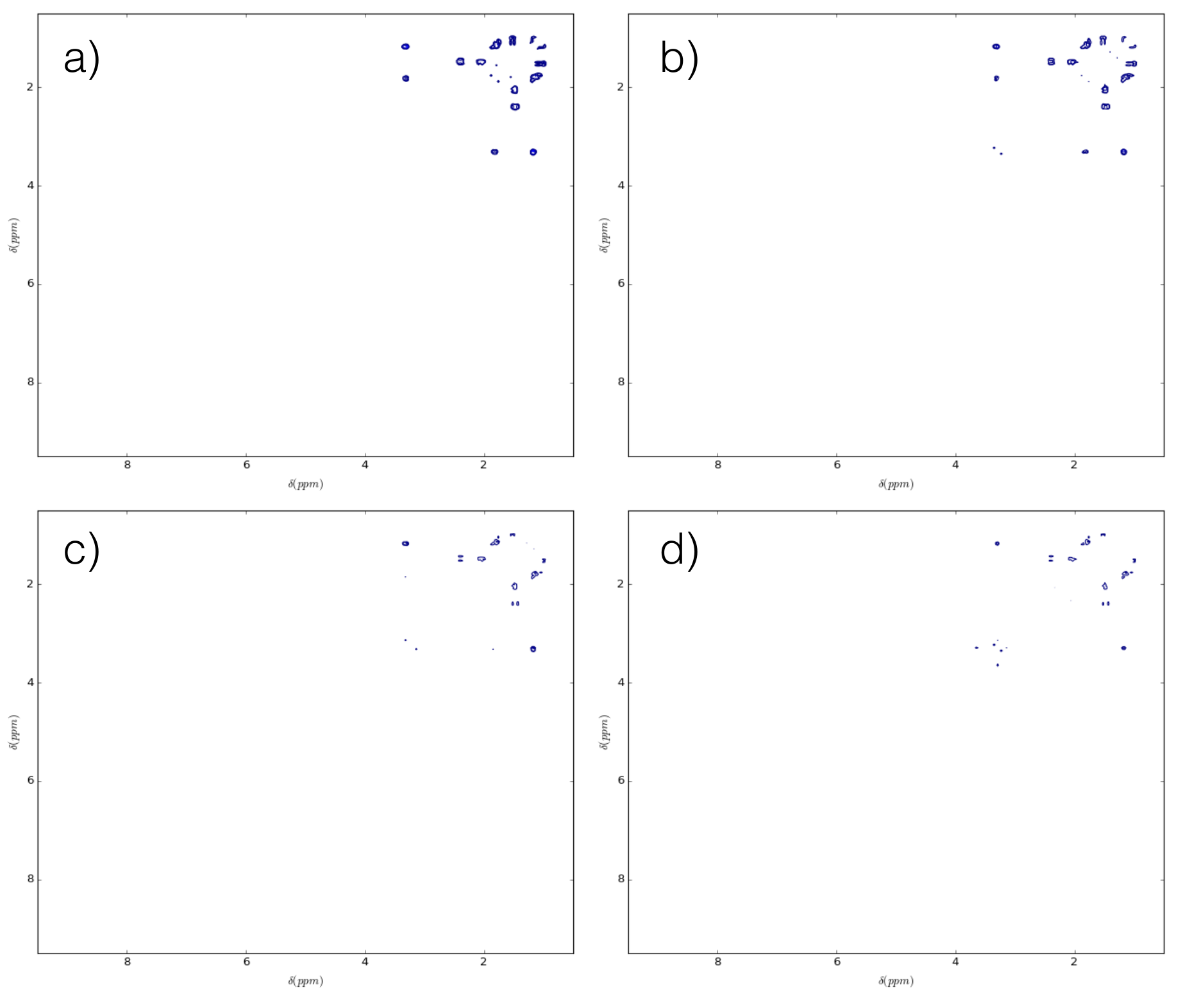}
\caption{Spectral fingerprint obtained by the ratio of bucket standard
deviation, a) comparison of sample 5 and sample 1, b-c-d) same as a for
sample 4, 3 and 2 respectively.\label{fig:Ratio}}
\end{figure}

The procedure above is not very sensitive, and the samples with lower
level of added artemisin could not be processed efficiently. A second
procedure was tested by taking the ratio of the bucket standard
deviation values. The results are shown in Figure \ref{fig:Ratio}:
despite a low level of concentration (few hundred micrograms of
artemisin in 10 mg of crude material), the spectral fingerprint is
recovered. In this case the diagonal of the homonuclear spectrum is not
recovered, this does not have a strong imapct, as it can be fully
infered from the off-diagonal fingerprint dots.

\subsubsection{Linear regression}\label{linear-regression}

This first approach take spectra two by two, and can be used on
homonuclear spectra, as shown here, and also on heteronuclear ones.
Using the whole set of spectra at once requires to have an additional
information, eventually imprecise, on the amount of active material in
each sample. In this case, signals coming from the studied molecule are
expected to be proportional to its concentration, and this property can
be exploited to separate those signals, varying along with the
concentration value, to the other signal, uncorrelated with it. This was
performed by using the \texttt{scikit-learn}
library\textsuperscript{24}, a generic tool for machine learning,
written in python with full interoperability with python, Jupyter and
SPIKE. The \texttt{linear\_model.LinearRegression()} function and the
Recursive Feature Elimination tool were used, and applied on the the
bucket lists area values (see S.I. for detailed operation).

\begin{figure}
\centering
\includegraphics[width=0.90000\textwidth]{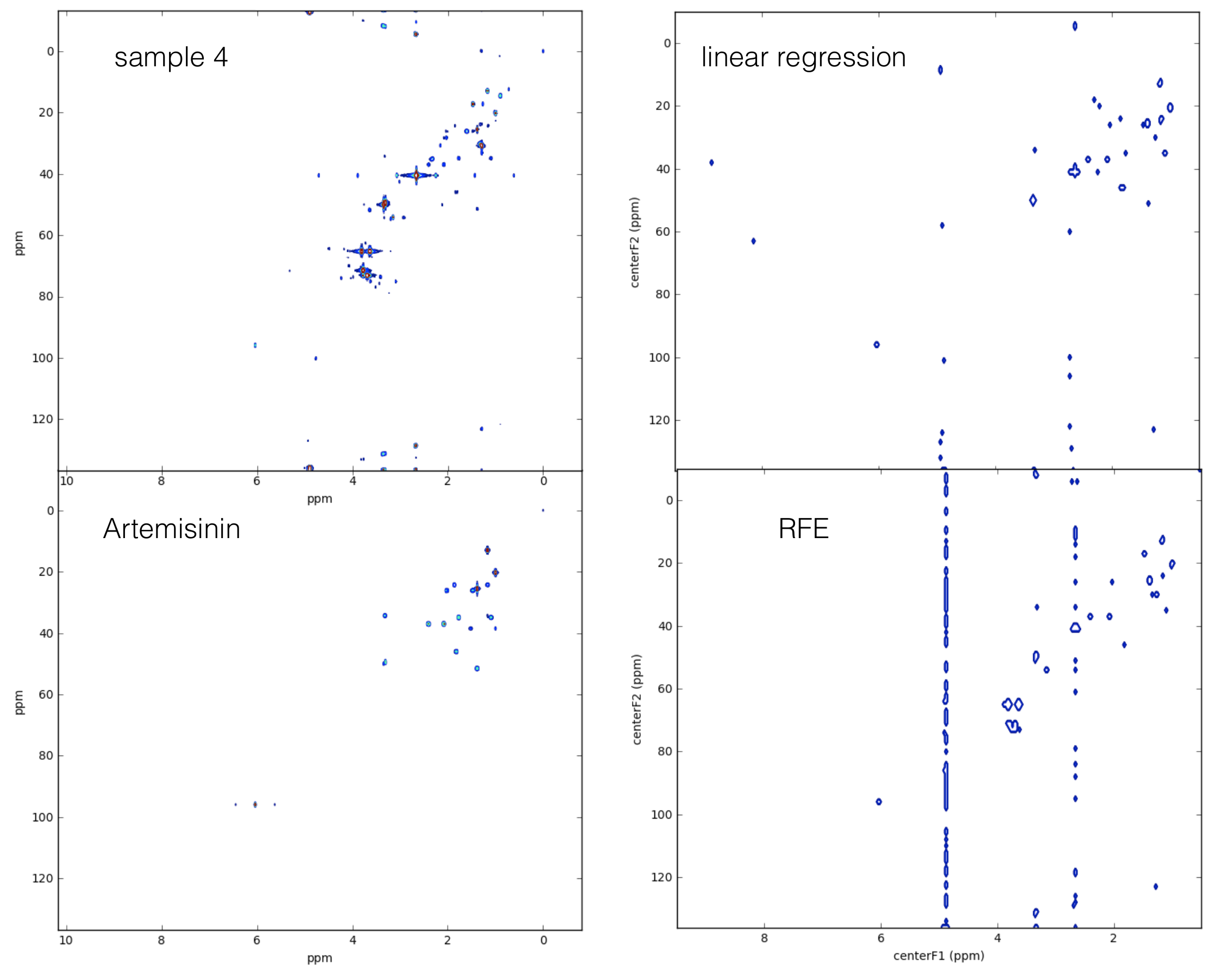}
\caption{HSQC Spectral fingerprint determined from the linear regression
analysis over the 4 samples with lowest
concentration.\label{fig:linear}}
\end{figure}

These tools allow to select a small subset of parameters which best
correlate with the estimated concentration. The selected features are
then supposed to define a spectral fingerprint in a manner equivalent
with the previous approach, but with a quantitative aspect this time. As
the whole spectrum series is used, it is expected to produce better
results. Results are shown in Figure \ref{fig:linear} for the HSQC
spectra obtained on the 4 samples presenting the lowest concentration.
It can be seen that the HSQC spectrum of artemisinin is extracted from
the complex spectrum of the mixture. The main artifacts observed in the
finger print are associated with the solvent lines (here water, methanol
and DMSO)

\section{Discussion}\label{discussion}

The measurement of NMR spectra over series of complex samples, and their
analysis, is a common procedure in screening or activity studies. The
acquisition part is usually well covered through the use of sample
changers and associated softwares, with the eventual help of companion
programs, allowing an optimized set-up\textsuperscript{25}. Here, we
extend the set of tools for these studies to the possibility of
automatic processing and statistical analysis of the set of 1D and 2D
spectra. There are already basic tools which allow to perform the first
processing steps of the data, however, they usually rely either on
preset parameters values (phase corrections, window function) or crude
estimate of the optimum parameters (baseline correction). In contrast,
the program Plasmodesma presented here, works in an autonomous manner,
without any user interaction, relying on a small set of preset global
parameters. It is able to autonomously process 1D, 2D, and DOSY
experiments, processing parameters are optimized either from the
experiment types (window function) or optimized automatically on the
data (phase correction, baseline). In addition, advanced methods are
used for the denoising of 2D spectra or the analysis of DOSY
experiments. Finally, the program generates spectra and peak lists and
bucket lists for all spectra, as well as reports on the data and the
analysis.

The use of the SPIKE library\textsuperscript{10}, a generalist
processing library for NMR and other Fourier spectroscopies, allowed a
rapid development of the program, as well as the use of advance tools.
Not relying on parameters previously by an operator allow to process
directly after the measure, and use the result of the processing as a
token for the quality of the acquisition. The processing step being more
rapid than the acquisition, it is perfectly possible to repeat the
processing along the series of measurements, to monitor the advance and
quality of the current experiment.

The series of spectra are further algorithmically analyzed using
machine-learning inspired approaches. However, each 2D spectrum is
typically composed of several million of data points, and this size
hampers the possibility to algorithmically compare efficiently several
spectra For instance the series of spectra generated in this study
represents more than 20 million points overall, and some
pre-conditioning of the data is required. For this reason, the automatic
analysis of the spectra is here principally performed on the bucket
lists, which provided a reduced but faithful representation of the
spectrum. Many artifacts such as antidiagonals, t\textsubscript{1} noise
and ridges, are also present. These artifacts are of rather low
intensity, however as the purpose here is the detection of compounds at
low concentration, their presence is detrimental, and we chose to
correct them on the bucket list rather than on the original spectra.
This smoothing and spectral normalization afforded by the bucketing
operation allows optimal spectral corrections and makes comparison
between spectra obtained from different samples easier. Finally, access
to quantities such as standard deviation of the signal, min and max
values, allows a finer description of the spectra.

From this material, a spectral fingerprint of the searched molecule
could be first determined from a two-by-two comparison of spectra with a
presence/absence of the searched compound, In this case the approach
consisting in comparing by ratio the standard deviation of buckets from
spectra of COSY type showed to be able to detect and extract the
spectral features of the compound even at low concentrations. Linear
regression over the whole series of spectra was also used to generate a
faithful fingerprint. One step regression as well as recursive feature
selection were used, and both proved to be efficient in extracting a
spectral fingerprint for both homonuclear and heteronuclear experiments
(see figure \ref{fig:linear} and S.I. S2).

Each experiment types can be used for the determination of the
fingerprint, and COSY type and HSQC type experiments were explored. It
is possible to perform the same analysis on a concatenation of all
experiments, however such an approach did not provide a correct result,
probably because of the heterogeneity of the different spectra types.

\section{Conclusion}\label{conclusion}

The program developed in this work represents an efficient alternative
for the autonomous processing of a series of NMR data (1D and 2D) and
contributes efficiently to the discovery of structurally unknown
molecules present in natural extracts, without any chromatographic
separation. It fully exploits the NMR technique as a fingerprinting
technique: complete 2D NMR fingerprint of the compound is recovered
through differential analysis performed both by comparison of local
variation in the spectra or by linear regression between signal
intensity and the concentrations of the natural product in the sample.

The extended bucketing procedure allows a strong reduction of the size
of the data, while preserving a large part of the molecular information
present in the original spectra. Basic machine learning approaches were
used to analyze this compressed but rich information, and proved
sufficient to readily extract the spectral fingerprint of the unknown
molecule, either from spectral comparison, or by handling of the whole
spectral series at once.

Plasmodesma is a novel tool which should be useful to decipher complex
mixtures, particularly in the discovery of biologically active natural
products from plants extracts, but can also in drug discovery or
metabolomics studies.

\section{Acknowledgments}\label{acknowledgments}

The authors are very grateful to Labex Medalis and Région Alsace for a
fellowship (LM), and Europe for an Erasmus fellowship (PM). We are also
grateful to A-M.Rusig for the collect and the identification of the
algal material, J.Viéville for help in the NMR set-up, and G.Bret for
help in the statistical analysis. We acknowledge
Wikimedia\textsuperscript{26} for the \emph{S.muticum} picture used in
the Graphical Abstract.

\section{E.S.I.}\label{e.s.i.}

\begin{itemize}
\tightlist
\item
  S1 Processing of the artemisinin series
\item
  S2 Analysis of the artemisinin series
\end{itemize}

\section*{References}\label{references}
\addcontentsline{toc}{section}{References}

\hypertarget{refs}{}
\hypertarget{ref-Bakiri2017}{}
{[}1{]} Bakiri, A.; Hubert, J.; Reynaud, R.; Lanthony, S.; Harakat, D.;
Renault, J.-H.; Nuzillard, J.-M. \emph{J Nat Prod.} \textbf{2017},
\emph{80} (5), 1387--1396.

\hypertarget{ref-Dabrosca2017}{}
{[}2{]} D'Abrosca, B.; Lavorgna, M.; Scognamiglio, M.; Russo, C.;
Graziani, V.; Piscitelli, C.; Fiorentino, A.; Isidori, M. \emph{Food
Chem Toxicol.} \textbf{2017}.

\hypertarget{ref-Abdelsalam2017}{}
{[}3{]} Abdelsalam, A.; Mahran, E.; Chowdhury, K.; Boroujerdi, A.;
El-Bakry, A. \emph{Physiol Mol Biol Plants.} \textbf{2017}, \emph{23}
(2), 369--383.

\hypertarget{ref-Hubert2014}{}
{[}4{]} Hubert, J.; Nuzillard, J.-M.; Purson, S.; Hamzaoui, M.; Borie,
N.; Reynaud, R.; Renault, J.-H. \emph{Anal Chem.} \textbf{2014},
\emph{86} (6), 2955--2962.

\hypertarget{ref-Oettl2014}{}
{[}5{]} Oettl, S.-K.; Hubert, J.; Nuzillard, J.-M.; Stuppner, H.;
Renault, J.-H.; Rollinger, J.-M. \emph{Anal Chim Acta.} \textbf{2014},
\emph{846}, 60--67.

\hypertarget{ref-Larive:2014vp}{}
{[}6{]} Larive, C. K.; Barding Jr, G. A.; Dinges, M. M. \emph{Anal Chem}
\textbf{2015}, \emph{87} (1), 133--146.

\hypertarget{ref-Mahrous2015}{}
{[}7{]} Mahrous, E.; Farag, M. \emph{J Adv Research} \textbf{2015},
\emph{6} (315).

\hypertarget{ref-Wolfender2015}{}
{[}8{]} Wolfender, J.-L.; Marti, G.; Thomas, A.; Bertrand, S. \emph{J
Chromatogr A.} \textbf{2015}, \emph{1382}, 136--164.

\hypertarget{ref-plasmo}{}
{[}9{]} Wikipedia. \emph{``Plasmodesma are microscopic channels which
traverse the cell walls of plant cells and some algal cells, enabling
transport and communication between them''}.
\emph{https://en.wikipedia.org/wiki/Plasmodesma}, {[}Online; accessed
16-May-2017{]}.

\hypertarget{ref-Spike2016}{}
{[}10{]} Chiron, L.; Coutouly, M.-A.; Starck, J.-P.; Rolando, C.;
Delsuc, M.-A. \emph{arXiv} \textbf{2016}.

\hypertarget{ref-Pons:1996wg}{}
{[}11{]} Pons, J. L.; Malliavin, T. E.; Delsuc, M.-A. \emph{J Biomol
NMR} \textbf{1996}, \emph{8}, 445--452.

\hypertarget{ref-Tramesel:2007ue}{}
{[}12{]} Tramesel, D.; Catherinot, V.; Delsuc, M.-A. \emph{J Magn Reson}
\textbf{2007}, \emph{188} (1), 56--67.

\hypertarget{ref-Pfandler:1987ve}{}
{[}13{]} Pfändler, P.; Bodenhausen, G.; Rapin, J.; Houriet, R.; Gäumann,
T. \emph{Chem Phys Let} \textbf{1987}, \emph{138} (2), 195--200.

\hypertarget{ref-Pfaendler:1988uo}{}
{[}14{]} Pfändler, P.; Bodenhausen, G.; Rapin, J.; Walser, M.; Gäumann,
T. \emph{J Am Chem Soc} \textbf{1988}, \emph{110} (17), 5625--5628.

\hypertarget{ref-Agthoven:2012cx}{}
{[}15{]} van Agthoven, M. A.; Delsuc, M.-A.; Bodenhausen, G.; Rolando,
C. \emph{Anal Bioanal Chem} \textbf{2013}, \emph{405}, 51--61.

\hypertarget{ref-Walt2011}{}
{[}16{]} van der Walt, S.; Colbert, S.; Varoquaux, G. \emph{Comput Sci
Eng} \textbf{2011}, \emph{13}, 22--30.

\hypertarget{ref-Jones2001}{}
{[}17{]} Jones, E.; Oliphant, T.; Peterson, P.; Others. SciPy: Open
source scientific tools for Python, 2001.

\hypertarget{ref-McKinney:2014tx}{}
{[}18{]} McKinney, W. \emph{Pandas, Python Data Analysis Library. 2015};
Reference Source, 2014.

\hypertarget{ref-hdf5}{}
{[}19{]} The HDF Group. Hierarchical Data Format, version 5.

\hypertarget{ref-Tables}{}
{[}20{]} Alted, F.; Vilata, I.; others. PyTables: Hierarchical datasets
in Python, 2002.

\hypertarget{ref-Chiron:2014hf}{}
{[}21{]} Chiron, L.; van Agthoven, M. A.; Kieffer, B.; Rolando, C.;
Delsuc, M.-A. \emph{Proc Natl Acad Sci USA} \textbf{2014}, \emph{111}
(4), 1385--1390.

\hypertarget{ref-Cherni:2017ds}{}
{[}22{]} Cherni, A.; Chouzenoux, E.; Delsuc, M.-A. \emph{Analyst}
\textbf{2017}, \emph{142} (5), 772--779.

\hypertarget{ref-Vonthron:2011}{}
{[}23{]} Vonthron-Sénécheau, C.; Kaiser, M.; Devambez, I.; Vastel, A.;
Mussio, I.; Rusig, A.-M. \emph{Mar. Drugs} \textbf{2011}, \emph{9},
922--933.

\hypertarget{ref-scikit-learn}{}
{[}24{]} Pedregosa, F.; Varoquaux, G.; Gramfort, A.; Michel, V.;
Thirion, B.; Grisel, O.; Blondel, M.; Prettenhofer, P.; Weiss, R.;
Dubourg, V.; Vanderplas, J.; Passos, A.; Cournapeau, D.; Brucher, M.;
Perrot, M.; Duchesnay, E. \emph{Journal of Machine Learning Research}
\textbf{2011}, \emph{12}, 2825--2830.

\hypertarget{ref-Clos:2013}{}
{[}25{]} Clos, L. J.; Jofre, M. F.; Ellinger, J. J.; Westler, W. M.;
Markley, J. L. \emph{Metabolomics} \textbf{2013}, \emph{9} (3),
558--563.

\hypertarget{ref-wiki:SMuticum}{}
{[}26{]} Commons, W. File:Sargassum muticum yendo fensholt 1955 lamiot
wimmereuxhautsdefrance estran juillet 2016a9.jpg --- wikimedia commons,
the free media repository, 2016.

\newpage
\section{Supp.Info 1} % (fold)
\includepdf[pages={-}]{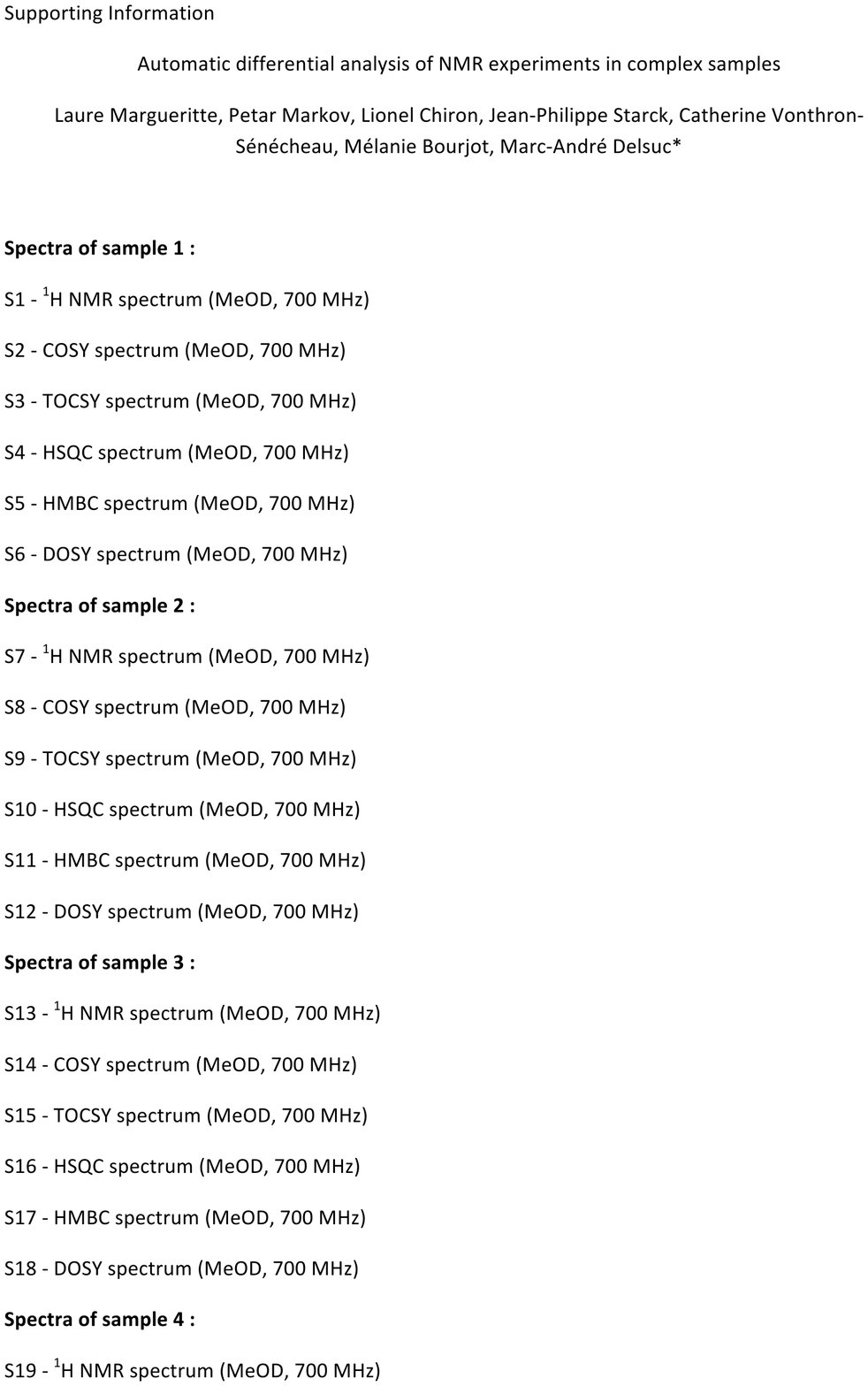}

\section{Supp.Info 2} % (fold)

Supp Info 2 can be found at https://github.com/delsuc/plasmodesma/blob/master/Analysis.ipynb

\end{document}